\documentclass[fleqn,prr,floatfix,showkeys,twocolumn]{revtex4}

\usepackage{graphicx}
\usepackage{amsmath}
\usepackage{dcolumn}
\usepackage{tabularx}
\usepackage[cp1252]{inputenc}
\usepackage{multirow}
\usepackage{booktabs}
\usepackage{placeins}
\usepackage{xcolor}
\begin{document}

\title{Density effects in precision laser spectroscopy of exotic helium atoms}%

\author{Hubert J. J\'{o}\'{z}wiak$^{1,2}$}
\email{hubert.jozwiak@umk.pl}
\author{Dimitar Bakalov$^3$}
\author{Micha\l{} Przybytek$^4$}
\author{Michail Stoilov$^3$}
\author{Piotr Wcis\l{}o$^1$}

\affiliation
{$^1$Institute of Physics, Faculty of Physics, Astronomy and Informatics, Nicolaus Copernicus University in Toru\'{n}, Grudziadzka 5, 87-100 Toru\'{n}, Poland}
\affiliation
{$^2$Institute for Molecules and Materials, Radboud University, Nijmegen, The Netherlands}
\affiliation
{$^3$Institute for Nuclear Research and Nuclear Energy, Bulgarian Academy of Sciences, 1040~Sofia, Bulgaria}
\affiliation
{$^4$University of Warsaw, Faculty of Chemistry, Pasteura 1, 02-093 Warsaw, Poland}

 \date{\today}

 \begin{abstract}
Exotic helium atoms act as unique atomic traps for heavy, negatively charged particles, protecting them from nuclear annihilation and nuclear capture on timescales long enough to enable high-precision laser spectroscopy. Such measurements serve as stringent tests of three-body quantum electrodynamics and offer a direct route to determining fundamental particle masses. Motivated by upcoming spectroscopic efforts targeting pionic ($\pi^{-\,4}\mathrm{He}^+$) and kaonic ($K^{-\,4}\mathrm{He}^+$) helium, we present a rigorous theoretical evaluation of the collisional and density effects governing these systems. Using an \textit{ab initio} potential energy surface and coupled-channel quantum scattering calculations, we study the collisional stability of the candidate metastable states against inelastic quenching in a cryogenic helium buffer gas. Furthermore, we provide theoretical reference values for the pressure broadening and pressure shift coefficients of the targeted transitions. These results establish an essential benchmark for future experiments, paving the way for refined determinations of the pion and kaon masses.
 \end{abstract}


\maketitle

\section{Introduction}
\begin{figure*}[!ht]
    \centering
    \includegraphics[width=\linewidth]{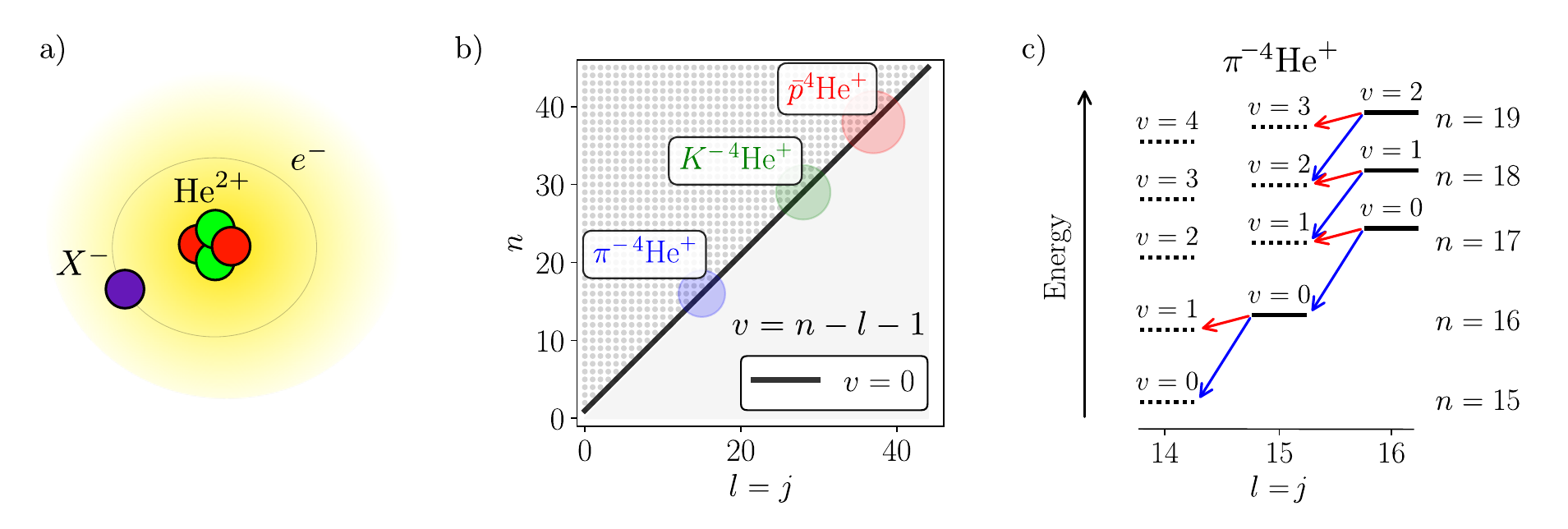}
    \caption{Physical structure, states, and transitions in exotic helium atoms. (a) Schematic representation of an exotic helium atom. A heavy, negatively charged particle $X^{-}$ ($X^- = \pi^{-}, K^{-}, \bar{p}$) is captured into a highly excited, near-circular orbit around the $\mathrm{He}^{2+}$ nucleus, while the remaining electron (gold cloud) occupies the $1s$ electronic orbital. (b) Overview of the state space in exotic atoms illustrating the exact correspondence between the atomic principal and orbital quantum numbers $(n,l)$ and the molecular vibrational and rotational quantum numbers $(v,j)$. Each grey dot represents a single $(n,l)$ [$(v,j)$] state. States sharing vibrational quantum number are aligned along the diagonal, with the $v=0$ line highlighted explicitly as an example. The colored shaded circles are schematic, indicating the general regions of the initial capture orbitals and the resulting islands of metastability for $\pi^{-}\mathrm{He}^+$, $K^{-}\mathrm{He}^+$, and $\bar{p}\mathrm{He}^+$, rather than their exact boundaries. (c) Representative energy-level diagram for pionic helium-4 ($\pi^{-\,4}\mathrm{He}^+$) demonstrating the hierarchical grouping of states into $n$-manifolds. Solid black lines denote metastable states, whereas dashed lines indicate states subject to rapid Auger decay. Arrows indicate examples of favored ($\Delta n = \Delta l = -1$, blue) and unfavored ($\Delta n \neq \Delta l = -1$, red) radiative transitions.}
    \label{fig1}
\end{figure*}

 The precision spectroscopy of exotic atoms has emerged as a powerful tool for testing quantum electrodynamics (QED), fundamental symmetries, and determining the properties of elementary particles with unprecedented accuracy. Among these systems, exotic helium occupies a unique position~\cite{Yamazaki_2002, Hori_2024, Hayano_2012}. Formed when a negatively charged particle $X^-$ (such as a muon $\mu$, pion $\pi^-$, kaon $K^-$, or antiproton $\bar{p}$) is captured by a helium atom in a nonelastic collision, the $X^-$ particle ejects one of the electrons and becomes bound by the Coulomb attraction of the helium nucleus, as schematically depicted in Fig.~\ref{fig1}(a).
 
The captured particle initially occupies highly excited, near-circular orbitals with a binding energy comparable to that of the ejected electron, while the remaining electron occupies the lowest-energy electronic orbital. Such exotic systems combine features of two-electron atoms and diatomic molecules~\cite{Yamazaki_2002}. Accordingly, their bound states can be labeled using either the hydrogen-like principal and orbital quantum numbers $(n,l)$ of the $X^-$ particle, or by the vibrational and rotational molecular quantum numbers $(v,j)$, connected by the relations $v=n-l-1$ and $j = l$. The principal quantum number of the initial capture orbital is approximately $n(X^-)=\sqrt{\mu_{X^-{\rm He}^{+}}/m_e}$, where $\mu_{X^-{\rm He}^{+}}$ is the reduced mass of the $X^{-}$He$^{+}$ system, and the orbital quantum number typically takes the largest possible value, $l=n-1$. In the molecular description, the initial orbital $(n,l=n-1)$ thus corresponds to the vibrational ground states, $(v=0, j=n-1)$. The distribution of these initial capture orbitals across the state space is illustrated in Fig.~\ref{fig1}(b).

The longevity of these states is governed by the competition between radiative decay, internal Auger transitions, and, at high helium bath density, collisional quenching. While radiative decay rates for favored transitions [$(n,l)\to(n-1,l-1)$ or $(v,j)\to(v,j-1)$] scale approximately as $\lambda_{\rm rad} \approx 3.4\times 10^{10}/n^3$ s$^{-1}$~\cite{Yamazaki_2002}, Auger transitions are highly suppressed for large multipolarities ($\Delta l$). Because the energy spacing between near-circular states in the capture region is significantly smaller than the ionization energy of helium ($\varepsilon_{\rm I}=24.59$ eV), Auger decay requires a large change in the principal quantum number, $\Delta n$, and consequently a large $\Delta l$. This deep suppression of the Auger channel results in metastable states with lifetimes on the order of microseconds---a phenomenon anticipated by Condo and Russel in the 1960s~\cite{Condo_1964, Russel_1969} and confirmed decades later~\cite{Yamazaki_1993}. These states form distinct ``islands of metastability'' bounded by the rapid Auger decay regions, as shown in Fig.~\ref{fig1}(b).

The microsecond lifetimes of these metastable states unlock the potential for high-precision laser spectroscopy. By stimulating a transition from a metastable state to a short-lived state, prompt de-excitation and subsequent annihilation or nuclear capture of the $X^-$ particle are induced, providing a nearly 100\% detection efficiency. This technique has allowed experimental determination of transition frequencies in antiprotonic helium with fractional accuracies approaching the parts-per-billion (ppb) level~\cite{Hori_2024}. From a theoretical standpoint, these three-body exotic helium atoms are highly appealing due to their relative simplicity and computational tractability. Transition frequencies in $\bar{p}\mathrm{He}^{+}$ have been evaluated by accounting for high-order QED, relativistic, and nuclear finite-size effects, as well as the interactions with surrounding helium target atoms, reaching sub-ppb levels of accuracy~\cite{Korobov_1999,Korobov_2009,Korobov_2014, Korobov_2014_bethe}. The comparison of these theoretical predictions with experimental results in antiprotonic helium yielded the most precise values of the antiproton-to-electron mass ratio and---at the time---the antiproton magnetic moment, alongside stringent tests of CPT invariance (see Refs.~\cite{Yamazaki_2002, Hayano_2012, Hori_2024} for a comprehensive review of the subject).

However, achieving fractional experimental accuracies at the parts-per-billion (ppb) level requires a rigorous accounting of systematic effects, particularly the density-dependent collisional shift and broadening of the spectral lines induced by the surrounding helium gas. While recent experiments with antiprotonic helium have managed to suppress this uncertainty by reducing the helium density by three orders of magnitude~\cite{Hori_2016}, this approach presents a significant trade-off: operating in such low-density environments inevitably leads to a loss of experimental signal. 

Alternatively, precision measurements can be performed at higher densities if the collisional effects are rigorously accounted for by fixing the density-dependent widths and shifts to accurate theoretical values. Indeed, \textit{ab initio} calculations of line-shape parameters in molecular hydrogen based on accurate H$_{2}$--H$_{2}$ potential energy surfaces (PES) allowed a significant reduction of systematic errors, aiding in the determination of transition frequencies and thus testing QED in two-electron systems~\cite{Wcislo_2018,Zaborowski_2020,Wojtewicz_2020,Lamperti_2023,Cygan_2025,Stankiewicz_2026}. In the context of exotic atoms, initial semi-classical evaluations of density effects were pioneered for antiprotonic helium~\cite{Bakalov_2000, Bakalov_2012}, though they relied on a PES for the exotic-helium--ordinary-helium system computed on a relatively sparse grid of \textit{ab initio} points. Recently, we provided a state-of-the-art PES for this system and performed rigorous coupled-channel calculations of pressure broadening and pressure shift coefficients for the 50 strongest transitions in antiprotonic helium-4~\cite{Jozwiak_2025}, establishing a robust benchmark for future experiments.

The landmark experiments in $\bar{p}\mathrm{He}^+$ have benefited from the favorable properties of this exotic atom: the stability of the antiproton, the existence of a relatively large number of metastable states with lifetimes in the microsecond range, and the ability to detect resonance conditions with near-perfect efficiency~\cite{Yamazaki_2002, Hori_2016}. However, most of these premises are not fulfilled for other exotic atoms, presenting severe theoretical and experimental challenges.

A primary challenge is the intrinsic instability of the mesons themselves. Unlike the stable antiproton, the free pion ($\pi^{-}$) and kaon ($K^{-}$) decay into lighter particles, with natural lifetimes of 26~ns and 12~ns, respectively~\cite{Workman_2022_review}, which impose an absolute upper limit on the longevity of the exotic atoms. 
Furthermore, pionic helium ($\pi^{-}\mathrm{He}^{+}$) lacks a large number of metastable states. Near the initial capture orbital ($n = 16$), only four states exhibit Auger decay rates slow enough for the atom to survive until the pion naturally decays~\cite{Hori_2014,Bai_2022}, as highlighted in Fig.~\ref{fig1}(c). Exploiting these few long-lived states, laser spectroscopy was successfully performed on the $(n=17,l=16)\to(n=17,l=15)$ transition~\cite{Hori_2020}, with plans to probe the $(n=17,l=16)\to(n=16,l=15)$ {and $(n=18,l=17)\to(n=17,l=16)$} transitions to improve the pion-to-electron mass ratio. To match the 4 ppb accuracy of recent QED calculations~\cite{Bai_2022}, the experimental resonance frequency must be corrected for collisional shifts that approach $\sim 100$~GHz at typical target densities. There are two available calculations of these density effects in $\pi^{-}\mathrm{He}^+$ based on older potential energy surfaces: one using a semiclassical approach, and another using numerical solutions of variable phase equations~\cite{Obreshkov_2016}. Alternatively, the density effects were also estimated using the long-distance asymptotics of the interatomic potential~\cite{Bai_2022}. The numerical results across these works diverge by up to 80\%. This massive discrepancy necessitates a fully quantum treatment of the problem on the new, refined potential energy surface of the exotic-helium--ordinary-helium system.

Similarly, the existence of metastable states in kaonic helium ($K^{-}\mathrm{He}^{+}$) has been experimentally confirmed~\cite{Yamazaki_1989}, with at least three states
exhibiting sufficiently slow Auger decay to permit laser spectroscopy~\cite{Aznabayev_2025}. Such experiments, currently in the planning stages~\cite{Soter_2022}, promise to significantly improve the $K^-$ mass accuracy. However, the density shift and broadening parameters for $K^-\mathrm{He}^{+}$ have never been evaluated.

Beyond line-shape parameters, the fundamental collisional stability of these candidate states remains an open question. While Auger decay rates determine the lifetime of an exotic atom in vacuum, precision measurements are performed in dense helium targets. In such environments, collisions with ordinary helium atoms pose two distinct problems: state-to-state inelastic quenching that could rapidly deplete the population of the state of interest, and nuclear capture of the $X^-$ particle by the nucleus of the ordinary helium atom, driven by purely attractive interactions at short range. To date, no rigorous \textit{ab initio} study has evaluated this overall collisional stability for pionic and kaonic helium, which is essential to establish the feasibility of high-precision spectroscopy in these systems.

In this work, we aim to close these gaps in the theory of exotic helium spectroscopy. Utilizing the fully quantum scattering framework and the highly accurate potential energy surface for a pair of exotic helium and ordinary helium atoms developed in Ref.~\cite{Jozwiak_2025}, we first systematically evaluate the collisional stability of the metastable states in pionic and kaonic helium. After identifying the collisionally robust states that survive both inelastic quenching and nuclear capture, we select the most promising candidate spectral lines and compute the pressure shift and broadening coefficients for these lines perturbed by atomic helium. Furthermore, we complete our previous analysis by providing state-of-the-art shift and broadening parameters for a set of antiprotonic helium-3 ($\bar{p}^3\mathrm{He}^+$) lines that have previously been studied experimentally.

\section{Selection of candidate transitions}

\begin{table*}[!t]
    \centering
    \caption{List of spectral transitions in the exotic helium atoms evaluated in this work. Transitions are categorized into favored $(\Delta n = \Delta l = -1)$ and unfavored $(\Delta n \neq \Delta l = -1)$ transitions. For antiprotonic helium-3, we focus on transitions that have been experimentally investigated. Auger lifetimes are derived from theoretical widths reported by Bai \textit{et al.}~\cite{Bai_2022} and Hori \textit{et al.}~\cite{Hori_2014} for pionic helium, Aznabayev \textit{et al.}~\cite{Aznabayev_2025} for kaonic helium,     and Korobov~\cite{Korobov_2014_bethe} for antiprotonic helium-3.}
    \label{tab:transition_list}
    \vspace{2mm}
    \begin{tabular}{llcccrr}
        \hline\hline
        \multirow{2}{*}{\textbf{Exotic atom}} & \multirow{2}{*}{\textbf{Transition type}} & \textbf{Atomic descriptors} & \textbf{Molecular descriptors} & \textbf{Wavelength} & \multicolumn{2}{c}{\textbf{Auger lifetimes}} \\
        \cmidrule(lr){6-7}
        & & $(n_{a},l_{a}) \to (n_{b},l_{b})$ & $(v_{a},j_{a}) \to (v_{b},j_{b})$ & $\lambda$ (nm) & $\tau_{a}$ (s) & $\tau_{b}$ (s) \\ \hline
         \multirow{8}{*}{$\pi^{-\,4}\mathrm{He}^{+}$} 
        & \multirow{5}{*}{Favored}   
         & $(16,15) \to (15,14)$ & $(0,15) \to (0,14)$ & $ 199.5$ & $ 5.93\times 10^{-8}$ & $2.35\times 10^{-12}$ \\
        & & $(17,16) \to (16,15)$ & $(0,16) \to (0,15)$ & $ 266.4$ & $ 1.1\times 10^{-4}$ & $ 5.93\times 10^{-8}$ \\
        & & $(18,16) \to (17,15)$ & $(1,16) \to (1,15)$ & $ 354.8$ & $ 8.64 \times 10^{-7}$ & $4.83\times 10^{-12}$ \\ 
        & & $(18,17) \to (17,16)$ & $(0,17) \to (0,16)$ & $ 362.9$ & \multicolumn{1}{c}{$-$} & $ 1.1\times 10^{-4}$ \\ 
        & & $(19,16) \to (18,15)$ & $(2,16) \to (2,15)$ & $ 462.6$ & $ 6.47\times 10^{-8}$ & $1.86\times 10^{-12}$ \\ 
        \cline{2-7}
        & \multirow{4}{*}{Unfavored} 
           & $(16,15) \to (16,14)$ & $(0,15) \to (1,14)$ & $ 1647.5$ & $ 5.93\times 10^{-8}$ & $2.5 \times 10^{-13}$ \\ 
        & & $(17,16) \to (17,15)$ & $(0,16) \to (1,15)$ & $ 1632.2$ & $ 1.1\times 10^{-4}$ & $4.83\times 10^{-12}$ \\
        & & $(18,16) \to (18,15)$ & $(1,16) \to (2,15)$ & $ 1977.5$ & $ 8.64\times 10^{-7}$ & $1.86\times 10^{-12}$ \\
        & & $(19,16) \to (19,15)$ & $(2,16) \to (3,15)$ & $ 2403.4$ & $ 6.47\times 10^{-8}$ & $4.76\times 10^{-13}$ \\
        \hline
        \multirow{6}{*}{$K^{-\,4}\mathrm{He}^{+}$}   
        & \multirow{3}{*}{Favored}   
          & $(26,25) \to (25,24)$ & $(0,25) \to (0,24)$ & $ 251.3$ & $2.81\times 10^{-8}$ & \multicolumn{1}{c}{$-$} \\
        & & $(28,26) \to (27,25)$ & $(1,26) \to (1,25)$ & $ 340.8$ & $1.85\times 10^{-8}$ & $1.52 \times 10^{-9}$ \\
        & & $(30,27) \to (29,26)$ & $(2,27) \to (2,26)$ & $ 465.9$ & $5.17\times 10^{-8}$ & $2.18 \times 10^{-9}$ \\
        \cline{2-7}
        & \multirow{3}{*}{Unfavored} 
          & $(26,25) \to (26,24)$ & $(0,25) \to (1,24)$ & $ 3114.8$ & $2.81\times 10^{-8}$ & $3.89 \times 10^{-9}$ \\ 
        & & $(28,26) \to (28,25)$ & $(1,26) \to (2,25)$ & $ 3062.5$ & $1.85\times 10^{-8}$ & $2.44\times 10^{-12}$ \\
        & & $(30,27) \to (30,26)$ & $(2,27) \to (3,26)$ & $ 3237.7$ & $5.17\times 10^{-8} $& $3.56 \times 10^{-9}$ \\
        \hline
        \multirow{6}{*}{$\bar{p}^{3}\mathrm{He}^{+}$} 
        & \multirow{4}{*}{Favored}
           & $(32,31) \to (31,30)$ & $(0,31) \to (0,30)$ & 287.4 & $2.34 \times 10^{-6}$ & $3.62  \times 10^{-9}$\\
        & & $(34,32) \to (33,31)$ & $(1,32) \to (1,31)$ & 364.4 & $7.08\times 10^{-7}$ & $1.46 \times 10^{-9}$ \\
        & & $(36,33) \to (35,32)$ & $(2,33) \to (2,32)$ & 463.9 & $4.15\times 10^{-8}$ & $1.48 \times 10^{-9}$ \\
        && $(38,34) \to (37,33)$ & $(3,34) \to (3,33)$ & 593.4 & $3.10\times 10^{-6}$ & $2.83 \times 10^{-9}$ \\
        \cline{2-7}
        & \multirow{1}{*}{Unfavored}
          & $(36,34) \to (37,33)$ & $(1,34) \to (3,33)$ & 723.9 & \multicolumn{1}{c}{$-$} & $2.83 \times 10^{-9}$ \\
        \hline\hline
    \end{tabular}
\end{table*}

The initial selection of viable transitions for precision laser spectroscopy is guided by two main factors: the isolated Auger lifetimes of the states involved ($\tau_{a}$ and $\tau_{b}$ for the transition $a\to b$) and the practical accessibility of the transition wavelengths. Table~\ref{tab:transition_list} lists the specific spectral lines in pionic, kaonic, and antiprotonic helium identified as potential experimental candidates based on these vacuum properties.

In pionic helium, Bai \textit{et al.}~\cite{Bai_2022} identified four metastable $(n,l)$ states: $(16,15)$, $(17,16)$, $(18,16)$, and $(19,16)$ with Auger lifetimes higher than $\tau \gtrsim 60\,\mathrm{ns}$, with the prominent $(17,16)$ lifetime reaching hundreds of microseconds. The electric dipole selection rules for favored transitions, $(n,l) \to (n-1,l-1)$, allow for only a single transition connecting two of these long-lived states: $(17,16) \to (16,15)$. Consequently, this transition represents the highest priority for experimental investigations, as Auger decay is deeply suppressed for both the initial and final states.

For the remaining three metastable states, the corresponding favored transitions---$(16,15) \to (15,14)$, $(18,16) \to (17,15)$, and $(19,16) \to (18,15)$---couple the metastable states to levels with significantly shorter Auger lifetimes ($\tau \sim 1$ ps), resulting in large natural linewidths. Furthermore, some of the favored transitions lie in the ultraviolet part of the spectrum, which is challenging to address with high-precision spectroscopy. An alternative strategy, successfully employed in the first experimental demonstration of $\pi^{-\,4}\mathrm{He}^{+}$ spectroscopy~\cite{Hori_2020}, is to probe the unfavored $(n,l) \to (n,l-1)$ transitions. While these transitions still couple the metastable and short-lived states, their transition wavelengths typically fall in the near-infrared range, making them accessible with standard tunable laser systems. 

{Guided by the upcoming experimental campaigns, we extend our primary candidate list with another favored transition: $(18,17) \to (17,16)$ \cite{hori_private}. We note that theoretical Auger decay rates for the initial $(18,17)$ state are currently not available in the literature. However, this transition remains an attractive target as it involves the long-lived $(17,16)$ state and lies at a favorable wavelength ($\lambda \approx 363\,\mathrm{nm}$).}
Therefore, our analysis for pionic helium covers 9 transitions in total: 5 favored $(n,l) \to (n-1,l-1)$ lines and 4 unfavored $(n,l) \to (n,l-1)$ transitions.

A similar selection strategy is applied to kaonic helium. Recent calculations identify the $(n,l) = (26,25), \,(28,26)$, and $(30,27)$ states as sufficiently stable for spectroscopy, with Auger lifetimes on the order of tens of nanoseconds~\cite{Aznabayev_2025}. The corresponding favored transitions couple these states to the $(25,24)$, $(27,25)$, and $(29,26)$ states, which are predicted to remain relatively stable ($\tau \sim 1$~ns). To provide a comprehensive dataset for future experiments, similarly to the pionic helium case, we consider both the favored, $(n,l) \to (n-1,l-1)$, and unfavored transitions, $(n,l) \to (n,l-1)$, as initial candidates.

We complement this list with 5 transitions in antiprotonic helium-3 ($\bar{p}^{3}\mathrm{He}^+$) that have been previously investigated experimentally.

While all the initial states listed in Table~\ref{tab:transition_list} exhibit sufficient lifetimes in a vacuum, their ultimate experimental viability depends on their survival in a dense helium target. Previous studies based on dynamic polarizabilities in kaonic helium have warned that some of these states may suffer from density-induced Stark quenching in dense media~\cite{Aznabayev_2025}, effectively reducing their lifetimes. To definitively resolve whether these specific states remain experimentally viable, in the following section, we subject all candidate states to a rigorous, fully quantum evaluation of their collisional stability.

\section{Collisional stability of the metastable states}

The collisional stability of the candidate states, as well as the eventual spectral line-shape parameters, is determined by the potential energy surface (PES) for a pair of exotic and ordinary helium atoms. We employ the highly accurate \textit{ab initio} $X^{-}\mathrm{He}^{+}$--$\mathrm{He}$ PES recently developed by our group for the study of antiprotonic helium-4~\cite{Jozwiak_2025}. The interaction energies were computed within the Born--Oppenheimer approximation using the supermolecular approach and the full configuration interaction method, with the results extrapolated to the complete basis set limit. 

The surface was evaluated over a grid of 26,505 geometries parametrized by Jacobi coordinates: the internuclear distance in the exotic atom, $r$, the distance between the center of mass of the exotic atom and the helium atom, $R$, and the $\theta$ angle between the axes defined by $r$ and $R$. Because the PES is computed within the Born--Oppenheimer approximation, it is universally applicable to systems isoelectronic with the $\bar{p}^4\mathrm{He}^+$--$^4$He system considered in Ref.~\cite{Jozwiak_2025}. To adapt the PES for collisions involving pionic helium, kaonic helium, or antiprotonic helium-3, we use a coordinate transformation that relates the coordinates of the new system $(R',r',\theta')$ to the original $(R,r,\theta)$ grid (see Appendix B in Ref.~\cite{Jozwiak_2025}).

\begin{figure*}
    \centering
    \includegraphics[width=0.9\linewidth]{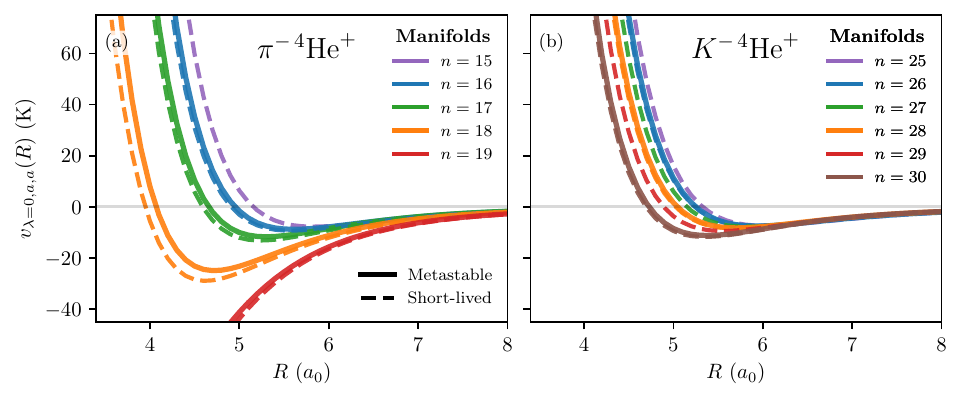}
    \caption{Effective isotropic expansion coefficients, $v_{\lambda=0,a,a}(R)$, for the candidate states in (a) pionic helium ($\pi^{-\,4}\mathrm{He}^+$) and (b) kaonic helium ($K^{-\,4}\mathrm{He}^+$) interacting with an ordinary helium atom. Curves are grouped and color-coded by the principal quantum number $n$ of the $X^{-}$ particle. Within each $n$-manifold, if present, solid lines denote the designated metastable state, while dashed lines represent the neighboring short-lived state participating in the transitions listed in Table~\ref{tab:transition_list}. The potential for the pionic $(18,17)$ state is omitted here to preserve visual clarity.}
    \label{fig:effective_potentials}
\end{figure*}

To rigorously study the scattering dynamics and determine the collisional stability of the candidate states, the 3D interaction potential must be expressed in the basis of the exotic atom's internal states. Following our previous work on quantum scattering in antiprotonic helium-4 \cite{Jozwiak_2025}, we choose to treat the $X^{-}\mathrm{He}^{+}$ exotic atom as a pseudo-diatomic molecule interacting with the incoming structureless helium atom, and we primarily adopt the molecular $(v,j)$ descriptors to describe the collision dynamics from this point forward. For quick reference, the exact correspondence between these molecular labels and the traditional atomic $(n,l)$ nomenclature for all states is provided in Table \ref{tab:transition_list}. As is standard in atom-diatom scattering theory, we expand the angular dependence of the PES in Legendre polynomials, $P_{\lambda}(\cos\theta)$,
\begin{equation}
\label{eq:potential-expansion}
    V(R, r, \theta) = \sum_{\lambda = 0} ^{\lambda_{\mathrm{max}}}A_{{\lambda}} (R, r) P_{\lambda} (\cos{\theta}),
\end{equation} 
where $\lambda_{\mathrm{max}}$ is determined by the relevance of each Legendre polynomial in representing the anisotropy of the $X^{-}\mathrm{He}^{+}$--$\mathrm{He}$ interaction potential. The dependence of the expansion coefficients, $A_{{\lambda}} (R, r) $, on the intramolecular exotic-particle--helium-nucleus distance, $r$, is projected by integrating over the wave functions, $\chi_{a}(r)$, of the isolated exotic helium atom
\begin{equation}
\label{eq:potential-rovibaverage}
    v_{{\lambda},a, a'} (R)= \int \mathrm{d}r\, \chi_{a}(r) A_{{\lambda}} (R, r) \chi_{a'}(r)  ,
\end{equation}
where $a \equiv (v_{a},j_{a})$ is a shorthand notation for the state of the exotic atom, and $v_{{\lambda},a, a'} (R)$ are the effective state-dependent expansion coefficients of the PES. These coefficients determine collisional dynamics. The diagonal terms ($a=a'$), specifically the isotropic term $\lambda = 0$, govern elastic scattering. The off-diagonal ($a \neq a'$) anisotropic ($\lambda \neq 0$) terms drive the inelastic state-to-state transitions.

Fig.~\ref{fig:effective_potentials} presents the effective isotropic expansion coefficients, $v_{\lambda=0,a,a}(R)$, computed for the candidate states in pionic and kaonic helium. We label and color-code these coefficients by the principal quantum number $n$ of the exotic particle. This visual representation is designed to map directly to the energy levels listed in Table \ref{tab:transition_list}. Specifically, within a given $n$-manifold, a solid line identifies an initial, long-lived metastable state $(v_a, j_a)$ [$(n_a, l_a)$]. The corresponding dashed line of the identical color represents the nearest-neighbor, short-lived state within that same manifold---the final state $(v_b, j_b)$ [$(n_b, l_b)$] in an unfavored transition. Note that manifolds lacking a metastable state are represented exclusively by dashed lines, indicating that only the short-lived states within that manifold are plotted.

The isotropic expansion coefficients within a given $n$-manifold exhibit similarities in their overall shape, well depth, and the position of their short-range repulsive walls, as their spatial extent is primarily governed by $n$. Comparing the two exotic atoms, $v_{\lambda=0,a,a}(R)$ for kaonic helium are notably shallower and exhibit repulsive walls pushed to larger internuclear distances ($R \sim 5.0\,a_0$) compared to the interactions involving pionic helium ($R \sim 4.0\,a_0$). We also note that the effective isotropic potential for the $(18,17)$ target state in pionic helium (omitted from Fig.~\ref{fig:effective_potentials} to preserve the visual hierarchy of the metastable states) exhibits this same standard repulsive wall at short range.

However, a significant difference emerges in the $v_{\lambda=0,a,a}(R)$ terms for the pionic helium $n=19$ manifold, corresponding to the vibrationally excited states $(v=2, j=16)$ and $(v=3, j=15)$. While the kaonic counterparts at similarly high vibrational levels (within manifolds $n=28, 29, 30$) maintain a standard attractive well and repulsive wall, the pionic $n=19$ isotropic terms are purely attractive. As established in prior studies of antiprotonic helium, collisional dynamics in these exotic systems are overwhelmingly elastic and dominated by the isotropic component of the interaction \cite{Bakalov_2000, Obreshkov_2004, Jozwiak_2025}. Consequently, $v_{\lambda=0,a,a}(R)$ serves, to a very good approximation, as an effective interaction potential for a given state. The complete absence of a repulsive wall in these terms is therefore a strong indicator of a loss of collisional stability.

This purely attractive behavior in pionic helium is a direct consequence of the different masses of the two exotic particles. Because the pion is significantly lighter than the kaon, its radial wave functions, $\chi_a(r)$, are much more spatially extended. In excited vibrational states like $v=2$, the pionic wave function probes configurations where the $\pi^-$ particle is actually closer to the incident ordinary helium nucleus than to its own He$^{2+}$ core. In such configurations, the incoming helium atom is exposed to an intense, unshielded Coulomb attraction that completely overwhelms the typical short-range exchange repulsion.

A non-reactive scattering formalism used in this work is incapable of describing the dynamics of such an attractive interaction potential. Physically, this means that the collision complex is drawn deep into the Coulomb well, leading to the subsequent capture of the pion by the ordinary helium nucleus.

To estimate the timescale of this process, we employ the classical Langevin capture model for barrierless reactions \cite{Clary_1990, Groenenboom_2010}. Because both the incident ordinary helium atom and the pionic helium atom are neutral species, the interaction is governed by the van der Waals dispersion potential. Fitting our \textit{ab initio} $n=19$ effective isotropic expansion coefficients reveals that the \textit{entire} attractive region is remarkably well-described by a $v_{\lambda = 0,a,a} = -C_6/R^6$.

In the classical framework of Langevin capture theory, reaction (here: capture of a pion by the ordinary helium nucleus) occurs with unit probability for collisions that possess sufficient kinetic energy $E$ to overcome the centrifugal barrier. For a $1/R^6$ potential, the critical capture cross-section is given analytically by
\begin{equation}
  \sigma(E) = 3\pi \left( \frac{C_6}{4E} \right)^{1/3}.  
\end{equation}
Averaging this cross-section over the Maxwell--Boltzmann distribution yields the temperature-dependent nuclear capture rate coefficient
\begin{equation}
  k_{\mathrm{capt}}(T) = 3\pi \sqrt{\frac{8}{\pi\mu}} \Gamma(5/3) \left( \frac{C_6}{4} \right)^{1/3} (k_B T)^{1/6}  ,
\end{equation}
where $\mu$ is the reduced mass of the collision system, $k_B$ is the Boltzmann constant, $T$ denotes temperature, and $\Gamma(x)$ is the gamma function.

Using the fitted $C_6$ coefficients for the $n=19$ manifold's effective isotropic expansion coefficients of the PES, this analytical expression yields $k_{\mathrm{capt}} \approx 1.5 - 2.2 \times 10^{-10}$~cm$^3$/s across the relevant experimental temperature range ($T = 1.5 - 15$ K). At the helium target densities used in Ref.~\cite{Hori_2020} ($\rho \sim 10^{22}$~cm$^{-3}$), this corresponds to a collisional lifetime on the order of picoseconds. Because this decay timescale is orders of magnitude shorter than the isolated Auger lifetimes of these states ($\tau > 10$~ns), high-precision spectroscopy is rendered impossible. Consequently, we exclude the $n=19$ manifold from further consideration, removing the $(v=2,j=16) \to (v=2,j=15)$ and $(v=2,j=16) \to (v=3,j=15)$ transitions from our list of viable experimental candidates. For all remaining states in both pionic and kaonic helium, the presence of a strong repulsive wall in the effective isotropic expansion coefficients suppresses this process, allowing us to rigorously evaluate their stability against inelastic quenching.

To visualize the most probable collisional paths for quenching of the metastable states, we refer back to the energy structure in pionic helium, shown schematically in Fig.~\ref{fig1}(c). Excluding the $n=19$ manifold---which we have established to be unstable against nuclear capture---the available energy levels closest to the metastable states lie within their own respective $n$ manifolds. The typical energy gaps between these adjacent $(v_a, j_a)$ [$(n_{a},l_{a})$] and $(v_a+1, j_a-1)$ [$(n_{a},l_{a}-1)$] states are on the order of $7\times 10^{3}\,\mathrm{K}$ in pionic helium and $4\times 10^{3}\,\mathrm{K}$ in kaonic helium. The energy intervals between adjacent $n$-manifolds are at least an order of magnitude larger. Because we are evaluating collisions in a helium buffer gas kept at cryogenic temperatures ($T \sim 1.5-15$ K), the available thermal kinetic energy is roughly three orders of magnitude too small to excite the atom to a higher state. Consequently, inelastic quenching is driven by the exothermic de-excitation to the nearest neighboring state lying just below the metastable level within the same $n$-manifold. This intra-manifold dominance was established earlier in antiprotonic helium \cite{Obreshkov_2004}, where the inter-manifold quenching is suppressed by five to seven orders of magnitude compared to transitions within the same $n$-manifold.

To evaluate the rate of this inelastic quenching from each $a \equiv (v_{a}, j_{a})$ metastable state to its nearest neighbor $a' \equiv (v_{a}+1, j_{a}-1)$ upon collision with an ordinary helium atom, we perform coupled-channel (CC) quantum scattering calculations on the \textit{ab initio} exotic-helium--ordinary-helium PES introduced above. The scattering wave function is expanded in a symmetry-adapted, body-fixed total angular momentum representation, yielding a set of second-order CC equations (see Ref.~\cite{Jozwiak_2025} for the explicit Hamiltonian and matrix elements). These equations are solved numerically using the renormalized Numerov algorithm~\cite{Johnson_1978} implemented in our in-house BIGOS scattering code~\cite{Jozwiak_2024a, Jozwiak_2024_code}. For computational details, see Appendix~\ref{app:A}.

\begin{figure*}[!ht]
    \centering
    \includegraphics[width=0.9\linewidth]{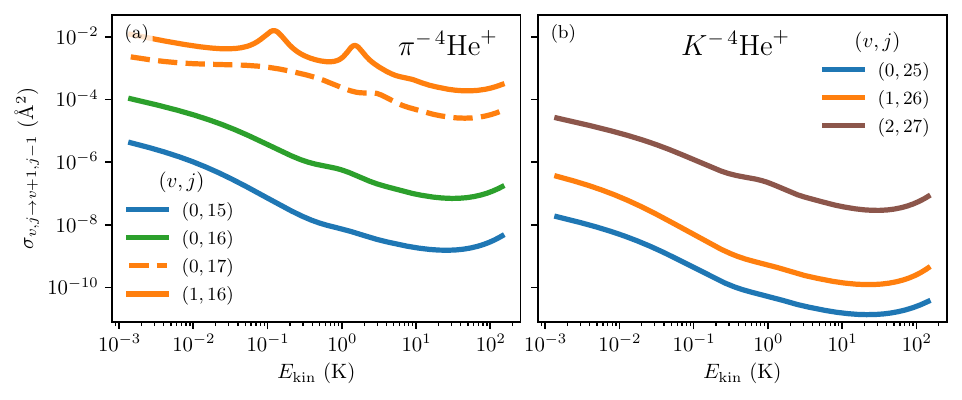}
    \caption{Leading inelastic quenching cross sections, $\sigma_{(v,j) \to (v+1, j-1)}$, originating from the three collisionally robust metastable states in (a) pionic helium and (b) kaonic helium. The color scheme matches the effective isotropic expansion coefficients of the PES presented in Fig.~\ref{fig:effective_potentials}, where the initial metastable state for each transition corresponds to the solid line of the identical color. For the pionic $n=18$ manifold (orange), we present cross-sections for both the metastable $(v=1,j=16)$ state (solid line) and the experimentally relevant $(v=0,j=17)$ state (dashed line), whose Auger lifetime is currently unavailable in the literature.}
    \label{fig:quench_xs}
\end{figure*}

At large intermolecular separations, the wave function is transformed to the space-fixed total angular momentum representation, and standard scattering boundary conditions are imposed to extract the scattering $S$-matrix. This energy-dependent $S$-matrix is then used to compute the state-to-state inelastic cross-section
\begin{align}
\begin{split}
&\sigma_{a \rightarrow a'}(E_{\mathrm{kin}}) =  \frac{\pi}{k^{2}(2j_{a}+1)}   \sum_{J, L, L'} (2J+1) \Bigl|S_{aL; a'L'}^{J}(E)\Bigr|^{2},
\end{split}
\end{align}
where $J$ is the total angular momentum of the collision complex, $L$ is the angular momentum of the relative motion, $E = E_{\mathrm{kin}} + E_{a}$ is the total energy, and $k = \hbar^{-1}\sqrt{2\mu E_{\mathrm{kin}}}$ is the initial wave vector.

The CC equations and cross-sections are evaluated over a dense logarithmic grid of kinetic energies spanning $E_{\mathrm{kin}} = 10^{-3}$ to $10^{2}\,\mathrm{K}$. This range is sufficient to rigorously thermally average the state-to-state cross-sections and extract the inelastic quenching rate coefficient,
\begin{align}
\begin{split}
k_{a \rightarrow a'}&(T) = \langle v_{r}\rangle   \int_{0}^{\infty} xe^{-x} \sigma_{a \rightarrow a'}(E_{\mathrm{kin}}=xk_{\mathrm{B}}T)  \,\mathrm{d}x,
\end{split}
\end{align}
where $\langle v_r \rangle = \sqrt{8k_{\mathrm{B}}T/(\pi\mu)}$ is the mean thermal relative velocity.

Fig.~\ref{fig:quench_xs} presents the state-to-state inelastic cross-sections for the metastable states of pionic (a) and kaonic (b) helium atoms considered in this work. In each case, based on the energy gap argument, the final state is $a' =(v_{a}+1, j_{a}-1)$. Although the Auger lifetime of the experimentally relevant pionic $(v=0,j=17)$ state is currently unknown, we evaluate the quenching cross-section for this state as well. Overall, the absolute magnitudes of these de-excitation cross-sections are extremely small, strictly bounded below $10^{-2}$~\AA$^2$ for the highest-rate process [pionic $(v=1,j=16)$] and dropping as low as $10^{-11}$~\AA$^2$ [kaonic $(v=0,j=25)$]. At the lowest collision energies, the cross-sections follow the Wigner threshold law for exothermic de-excitations, $\sigma \propto E_{\mathrm{kin}}^{-1/2}$. As the collision energy increases, the cross-sections pass through a minimum and rise toward the upper limit of the energy range considered. Additionally, the pionic $(v=1,j=16)$ cross-section exhibits two pronounced features at roughly $0.12$~K and $1.5$~K, which we attribute, via partial wave decomposition of the cross-sections, to the $p$ and $d$ partial wave scattering resonances, respectively.

Fundamentally, the overall smallness of these inelastic cross-sections is a direct consequence of the vast energy gap to the nearest kinematically allowed final state. However, the dramatic variations in magnitude---spanning several orders of magnitude between the two species and across different state manifolds---are dictated by the spatial extent of the exotic particle's wave function and its subsequent coupling to the anisotropic expansion terms of the potential [$\lambda \neq 0$ in Eq. \eqref{eq:potential-expansion}], which drive inelastic transitions. Because the pion is significantly lighter than the kaon, the radial wave functions of pionic helium extend to much larger interparticle distances. This increased spatial extent yields stronger effective anisotropic expansion terms [Eq.~\eqref{eq:potential-rovibaverage}], resulting in quenching cross-sections that are orders of magnitude larger than those of the more compact kaonic system. This exact same mechanism governs the hierarchy of states within a single species: states occupying higher $n$-manifolds or possessing greater vibrational excitation (e.g., $v=1, 2$) inherently feature broader spatial extents of the wave function, enhancing the inelastic quenching probability. A similar finding---that the inelastic quenching rate for collisional transitions with comparable energy gaps increases with $n$---was previously reported for antiprotonic helium in Ref. \cite{Obreshkov_2004}.

Thermally averaging these cross-sections over the experimental $1.5-15$ K temperature range yields quenching rate coefficients. The relatively smooth energy dependence of the cross-sections across the thermally relevant regime leads to a mild temperature dependence in the final rates. For pionic helium, the rates are on the order of $10^{-20}$, $10^{-18}$, $10^{-16}$ and $10^{-15}$~cm$^3$/s for the $(v=0,j=15)$, $(v=0,j=16)$, $(v=0,j=17)$, and $(v=1,j=16)$ states, respectively. For kaonic helium, the rates are even lower, ranging from roughly $10^{-22}$~cm$^3$/s for the $(v=0,j=25)$ state up to $10^{-19}$~cm$^3$/s for the $(v=2,j=27)$ state.

\begin{figure*}[!ht]
    \centering
    \includegraphics[width=0.9\linewidth]{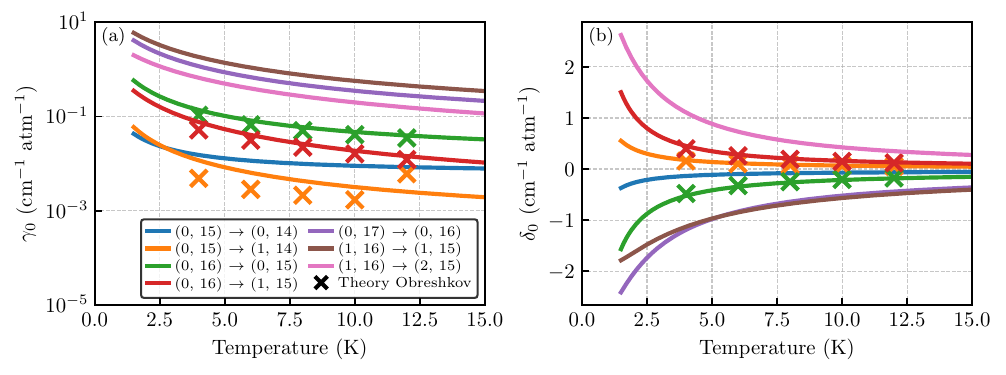}
    \caption{Pressure broadening ($\gamma_{0}$) and pressure shift ($\delta_{0}$) coefficients for the transitions in $\pi^{-\,4}\mathrm{He}^{+}$ considered in this work. The transitions are denoted by their molecular $(v_a, j_a) \to (v_b, j_b)$ descriptors; see Table~\ref{tab:transition_list} for their exact correspondence to the atomic $(n_a, l_a) \to (n_b, l_b)$ nomenclature. Markers denote previous theoretical results  ($\times$, Ref.~\cite{Obreshkov_2016}).}
    \label{fig:PiHe4}
\end{figure*}

To put these values into experimental perspective, we evaluate the collisional lifetimes [$\tau_{\mathrm{coll}} = 1/(\rho k)$] at the liquid-helium target density utilized in Ref.~\cite{Hori_2020} ($\rho = 2.18 \times 10^{22}$~cm$^{-3}$). For the state with the highest inelastic quenching rate---the $(v=1,j=16)$ metastable state in pionic helium---this extreme density yields a collisional lifetime of approximately $45$~ns. Comparing this to the isolated Auger lifetime of this state ($\tau \approx 864$~ns), it is evident that at such high target densities, collision-induced quenching would outpace the natural decay, significantly depleting the metastable population. However, at moderate gas-phase target densities (e.g., $\rho \sim 10^{20} - 10^{21}$~cm$^{-3}$), this collisional lifetime extends well into the microsecond regime, safely exceeding the natural Auger decay. For all remaining states across both species, the quenching lifetimes range from tens of microseconds to hundreds of milliseconds even at liquid-helium densities, rendering inelastic quenching entirely negligible.

Having evaluated the inelastic quenching rates and formally excluded the $n=19$ pionic manifold due to nuclear capture, we have definitively established the collisionally robust metastable states. We now turn to the primary objective of this work: evaluating the collisional line-shape parameters for the transitions involving these states.

\section{Pressure broadening and shift coefficients}

The theoretical framework for computing the collisional line-shape parameters of exotic helium atoms was established in our previous work~\cite{Jozwiak_2025}. While the inelastic quenching rates evaluated in the previous section depend only on the scattering dynamics of a single isolated state, the calculation of line-shape parameters for a spectroscopic transition $a \to b$ within the impact approximation requires the coherent combination of \textit{two} scattering $S$-matrices, evaluated at two distinct total energies: $E_{\mathrm{kin}} + E_{a}$ and $E_{\mathrm{kin}} + E_{b}$, where` $E_{\mathrm{kin}}$ is the relative kinetic energy of the collision, and $E_{a}$ and $E_{b}$ are the energies of the states coupled by the electromagnetic field. These two scattering $S$-matrices determine the complex generalized spectroscopic cross-section, $\sigma^{\kappa}(a, b; E_{\mathrm{kin}})$, whose real and imaginary components correspond to the pressure broadening and pressure shift, respectively, of the collision-perturbed $a \to b$ spectral transition of rank $\kappa$. In the space-fixed total angular momentum representation, this cross-section is given by \cite{BenReuven_1966, Shafer_1973}
\begin{equation}\label{eq:gsxs}
\begin{split}
&\sigma^{\kappa}(a, b; E_{\mathrm{kin}}) =
\frac{\pi}{k^2} \sum_{J_a,J_b,L,L^{'}} (-1)^{L+L^{'}}(2J_a+1)( 2J_b+1) \\&\times
\begin{Bmatrix}
j_b & \kappa & j_{a}\\
J_a & L & J_{b}
\end{Bmatrix}
\begin{Bmatrix}
j_b & \kappa & j_{a}\\
J_a & L' & J_{b}
\end{Bmatrix}
\\
&\times \Bigl(\delta_{LL^{'}} - S^{J_{a}}_{a L^{'};a L}(E_{\mathrm{kin}}+E_{a}) S^{J_{b}\,*}_{b {L}^{'};b {L}}(E_{\mathrm{kin}}+E_{b})\Bigr),
\end{split}
\end{equation}
where the symbols in braces are the Wigner 6-$j$ symbols, and $\kappa$ is the tensorial rank of the radiation-matter interaction. For the electric dipole transitions considered here, $\kappa = 1$. 

The thermal average of this cross-section yields the line-shape parameters: the pressure broadening ($\gamma_{0}$) and pressure shift ($\delta_{0}$) coefficients \cite{Wcislo_2021}:
\begin{equation}
    \gamma_{0} - i\delta_{0} = \frac{1}{2\pi c} \frac{\langle v_{r}\rangle}{k_{\mathrm{B}}T} \int_{0}^{\infty} x e^{-x} \sigma^{\kappa}(a,b; E_{\mathrm{kin}}=x k_{\mathrm{B}}T) \,\mathrm{d}x,
\end{equation}
where $c$ is the speed of light in vacuum. 
    
\subsection{Line shape parameters in pionic helium}

Having formally excluded the $n=19$ manifold, we focus on the four collisionally robust metastable states in pionic helium: $(v=0,j=15)$, $(v=0,j=16)$, $(v=0,j=17)$, and $(v=1,j=16)$. As listed in Table~\ref{tab:transition_list}, we evaluate 7 viable spectral lines involving these states: 4 favored $(\Delta v= 0, \Delta j=-1 )$ [$(\Delta n = \Delta l  = -1)$] and 3 unfavored $(\Delta v= +1, \Delta j=-1 )$ [$(\Delta n = 0, \Delta l=-1)$] transitions.

Fig.~\ref{fig:PiHe4} presents the computed pressure broadening ($\gamma_{0}$) and pressure shift ($\delta_{0}$) coefficients for these transitions across the relevant experimental temperature range of $1 - 15$ K. Both quantities exhibit a smooth, monotonically decreasing behavior with temperature. This reflects the gradual energy dependence of the generalized cross-sections, $\sigma^{\kappa}$. For transitions originating from the shallower $(v=0,j=15)$ and $(v=0,j=16)$ effective isotropic expansion coefficients [$n=16$ and $n=17$ manifolds in Fig.~\ref{fig:effective_potentials}(a)], $\sigma^{\kappa}$ is entirely featureless. While the deeper effective potential of the $(v=1,j=16)$ state ($n=18$) do support scattering resonances (discussed in Appendix~\ref{app:B}), thermal averaging over the Maxwell--Boltzmann distribution washes out these features, resulting only in a slight overall elevation of the line-shape parameters' magnitude.

Consistent with our previous findings for antiprotonic helium-4~\cite{Jozwiak_2025}, we observe that the pressure shift coefficients are systematically on the order of, or significantly larger than, the corresponding pressure broadening coefficients. This behavior is characteristic of light molecules with sparse, widely separated energy levels where collision-induced inelastic transitions are highly suppressed \cite{Wcislo_2021,Stankiewicz2021}. As established in the previous section, the energy gap on the order of $10^{3}\,\mathrm{K}$ to the nearest asymptotically open channel ensures that the broadening is driven almost exclusively by elastic dephasing---the accumulation of phase differences between the elastic scattering amplitudes of the initial and final spectroscopic states during a collision.

\begin{figure*}[!ht]
    \centering
    \includegraphics[width=0.9\linewidth]{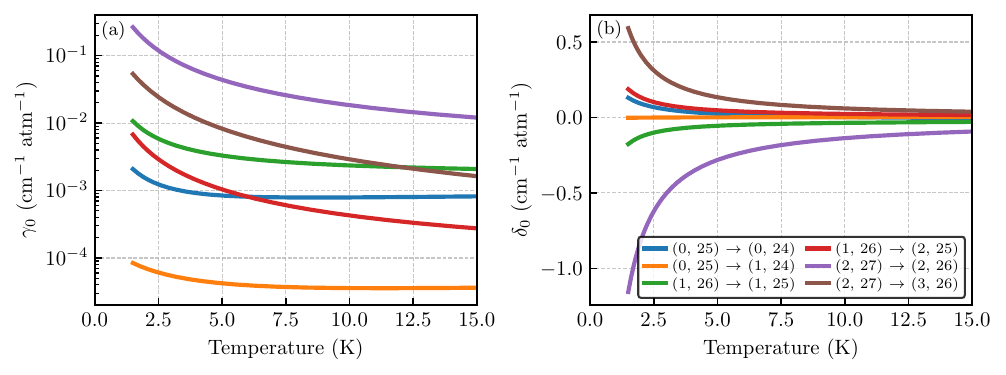}
    \caption{Pressure broadening ($\gamma_{0}$) and pressure shift ($\delta_{0}$) coefficients for the transitions in $K^{-\,4}\mathrm{He}^{+}$ considered in this work. The transitions are denoted by their molecular $(v_a, j_a) \to (v_b, j_b)$ descriptors; see Table~\ref{tab:transition_list} for their exact correspondence to the atomic $(n_a, l_a) \to (n_b, l_b)$ nomenclature.}
    \label{fig:KHe4}
\end{figure*}

Because this elastic dephasing, to a good approximation, is governed by the difference between the effective isotropic expansion coefficients of the PES for the initial and final states, $\Delta v_{\lambda=0} = v_{\lambda=0,a,a} - v_{\lambda=0,b,b}$, the line-shape parameters in Fig.~\ref{fig:PiHe4} reveal a strict physical hierarchy.

The first level of this hierarchy distinguishes the \textit{type} of transition. For any given metastable state, the favored transition $(\Delta v = 0, \Delta j=-1)$ [$(\Delta n = \Delta l = -1)$] exhibits a significantly larger pressure broadening and absolute pressure shift than the corresponding unfavored transition $(\Delta v = +1, \Delta j=-1)$ [$(\Delta n = 0, \Delta l=-1)$]. This is a direct reflection of the shape of the effective isotropic terms shown in Fig.~\ref{fig:effective_potentials}(a): an unfavored transition to a state within the same $n$-manifold involves a slight deepening of the effective potential, whereas a favored transition to a lower $n$-manifold involves jumping to a significantly shallower potential. Furthermore, the shifts for favored and unfavored transitions originating from the same state possess opposite signs, confirming that the perturbation is driven by whether the final state potential is more or less repulsive than the initial one.

The second level of the hierarchy distinguishes the \textit{initial state}. The magnitude of the differential potential, $\Delta v_{\lambda=0}$, grows rapidly with the principal quantum number $n$. Consequently, transitions originating from the higher-lying $(v=1,j=16)$ state ($n=18$) exhibit the largest absolute shift and broadening coefficients, while those originating from the $(v=0,j=15)$ state ($n=16$) are the least perturbed by the helium bath. 

{Among all considered candidates, there is only one instance where two transitions connect the identical pair of $n$-manifolds: the favored $(v=1,j=16) \to (v=1,j=15)$ and $(v=0,j=17) \to (v=0,j=16)$ lines, both representing $n=18 \to 17$ transitions. Because the spatial extent of the exotic particle's wave function is governed primarily by $n$, the corresponding differential isotropic potentials are similar, and, consequently, their pressure broadening and shift coefficients closely match.}

Fig.~\ref{fig:PiHe4} also compares our results with the previous theoretical calculations by Obreshkov \textit{et al.}~\cite{Obreshkov_2016}. To ensure a direct comparison, we plot their calculations derived using a coupled partial-wave approach on the ``HN1'' fit of the \textit{ab initio} points (see Ref. \cite{Bakalov_2000} for details)---a methodology conceptually equivalent to our own. We find overall excellent agreement with these results. The only notable discrepancy occurs in the pressure broadening of the $(v=0,j=15) \to (v=1,j=14)$ transition. We attribute this deviation to the refined \textit{ab initio} grid of the new PES~\cite{Jozwiak_2025}, which more accurately captures the molecular configurations relevant to the excited $(v=1,j=14)$ state. This discrepancy highlights the importance of our new \textit{ab initio} calculations as a rigorous theoretical benchmark for future precision metrology in pionic helium.

Finally, we consider the first experimental detection of pionic helium~\cite{Hori_2020}, which probed the $(v=0,j=16) \to (v=1,j=15)$ transition in a superfluid helium target at 4~K. While our binary collision framework is strictly valid only for dilute gas phases, evaluating our results at the experimental liquid density ($\rho = 2.18 \times 10^{22}$~cm$^{-3}$) provides an order-of-magnitude estimate for the perturbation. Using our calculated parameters at 4~K ($\gamma_{0} = 7.6 \times 10^{-2}$~cm$^{-1}$ atm$^{-1}$ and $\delta_{0} = 0.45$~cm$^{-1}$ atm$^{-1}$), we estimate a collisional linewidth of $\sim 27$~GHz and a shift of $\sim 161$~GHz. We note that the experimentally observed line center was shifted by roughly $78$~GHz toward higher frequencies, demonstrating the necessity of accounting for dense-medium effects beyond the binary collision approximation in future liquid-phase experiments.

\subsection{Line shape parameters in kaonic helium}

Fig.~\ref{fig:KHe4} presents the computed pressure broadening and shift coefficients for the 3 favored $(\Delta v= 0, \Delta j=-1 )$ [$(\Delta n = \Delta l  = -1)$] and 3 unfavored $(\Delta v= +1, \Delta j=-1 )$ [$(\Delta n = 0, \Delta l=-1)$] transitions in the kaonic helium atom. To our knowledge, these are the first rigorous theoretical calculations of collisional line-shape parameters for the kaonic helium atom. While there is currently no experimental or theoretical data available for direct comparison, we can quantitatively contrast these results with our pionic helium calculations.

Overall, both the pressure broadening and shift coefficients in kaonic helium are significantly smaller in magnitude. This is a direct consequence of the characteristics of the effective isotropic expansion coefficients shown in Fig.~\ref{fig:effective_potentials}. Because the kaon is significantly heavier than the pion, the radial wave functions of the kaonic helium atom are much more tightly localized. Consequently, integrating the identical underlying Born--Oppenheimer potential energy surface over these narrow wave functions probes a much smaller range of the exotic atom's internal coordinates. This results in effective isotropic expansion coefficients that are overall shallower, and whose well depths vary only by roughly 50\% across the states of interest. Consequently, the differential potentials ($\Delta v_{\lambda=0}$) between the initial and final spectroscopic states are substantially reduced, leading to weaker elastic dephasing. This is most strikingly illustrated by the $(v=0,j=25) \to (v=1,j=24)$ unfavored transition; the effective interaction potentials for these two states belonging to the $n=26$ manifold are visually indistinguishable in Fig.~\ref{fig:effective_potentials}(b), resulting in exceptionally small line-shape parameters ($\gamma_0 \lesssim 10^{-4}$~cm$^{-1}$ atm$^{-1}$ and $|\delta_0| < 10^{-3}$~cm$^{-1}$ atm$^{-1}$).

\begin{figure*}[!ht]
    \centering
    \includegraphics[width=0.9\linewidth]{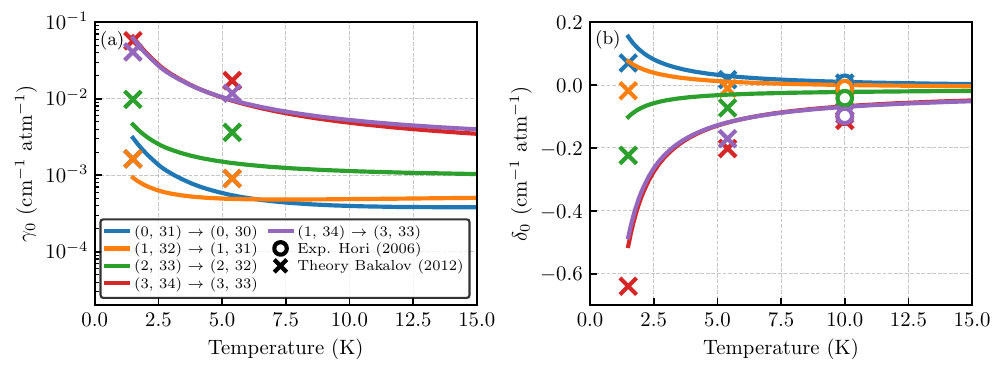}
    \caption{Pressure broadening ($\gamma_{0}$) and pressure shift ($\delta_{0}$) coefficients for the transitions in $\bar{p}^{3}\mathrm{He}^{+}$ considered in this work. The transitions are denoted by their molecular $(v_a, j_a) \to (v_b, j_b)$ descriptors; see Table~\ref{tab:transition_list} for their exact correspondence to the atomic $(n_a, l_a) \to (n_b, l_b)$ nomenclature.  Markers denote previous theoretical ($\times$, Ref. \cite{Bakalov_2012}) and experimental ($\circ$, Ref. \cite{Hori_2006}) results.}
    \label{fig:apHe3}
\end{figure*}

Despite these small absolute magnitudes, the qualitative behavior of the kaonic helium line-shape parameters mirrors the trends established for pionic helium. Both coefficients exhibit a smooth, monotonic decrease with temperature, and the pressure shifts strictly dominate the corresponding broadenings due to the suppression of inelastic quenching.

Furthermore, the two-fold physical hierarchy remains fully intact. First, favored transitions $(\Delta v = 0, \Delta j=-1)$ [$(\Delta n = \Delta l = -1)$] consistently exhibit larger pressure broadenings than their unfavored counterparts $(\Delta v = +1, \Delta j=-1)$ [$(\Delta n = 0, \Delta l=-1)$] originating from the same metastable state, and their corresponding shifts carry opposite signs. Second, transitions originating from higher-lying states, such as $(v=2,j=27)$ ($n=30$), experience stronger density effects than those originating from lower-lying states like $(v=1,j=26)$ ($n=28$). In all cases, these hierarchies are determined by the identical physical mechanism established previously: the magnitude of the differential effective interaction potential during the collision.

Finally, as detailed in Appendix B, the overall shallowness of the $K^{-}\mathrm{He}^{+}$--He effective interaction potentials does not allow for the existence of any bound states. Consequently, the generalized spectroscopic cross-sections do not exhibit the sharp resonant features observed in the pionic system. While mild cross-section enhancements arise from the temporary trapping of specific partial waves ($L = 1$), these effects are smoothly integrated out during thermal averaging, yielding the strictly featureless temperature dependencies observed in Fig.~\ref{fig:KHe4}.

\subsection{Line shape parameters in antiprotonic helium-3}

We complete our survey of exotic helium atoms by presenting the reference line-shape parameters for antiprotonic helium-3 ($\bar{p}^3\mathrm{He}^+$) perturbed by ordinary $^3\mathrm{He}$ buffer gas. Table~\ref{tab:transition_list} lists the five transitions evaluated for this system. Four of these are favored transitions $(\Delta v = 0, \Delta j=-1)$ [$(\Delta n = \Delta l = -1)$]. The single remaining transition is unfavored, which in this case corresponds to  $(\Delta v = +2, \Delta j=-1)$ [$(\Delta n = +1, \Delta l = -1)$]. The computed pressure broadening and pressure shift coefficients are presented in Fig.~\ref{fig:apHe3}.

Consistent with the behavior established in pionic and kaonic helium, the pressure shift coefficients for all five transitions are significantly larger than their corresponding pressure broadening coefficients. As previously discussed, this confirms that inelastic quenching is highly suppressed by the large energy gaps to adjacent states, leaving elastic dephasing as the dominant broadening mechanism. Furthermore, the parameters for all transitions exhibit the expected monotonic decrease in magnitude as the temperature increases across the $1 - 15$~K range.

Pressure broadening strictly follows the established hierarchy, growing with increasing principal quantum number $n$. The pressure shift coefficients for the low-$n$ transitions, $(v=0,j=31) \to (v=0,j=30)$ and $(v=1,j=32) \to (v=1,j=31)$, are positive. For transitions higher up the energy ladder, the shift reverses sign (becoming negative), and its absolute magnitude grows monotonically with $n$. This sign reversal reflects a crossover in the differential effective potential, $\Delta v_{\lambda=0}$; for the lowest states, the final-state potential is more repulsive than the initial state, while for higher $n$, this trend inverts. 

Finally, Fig.~\ref{fig:apHe3} compares our results with the previous theoretical calculations by Bakalov \textit{et al.}~\cite{Bakalov_2012} and the 10~K experimental measurements by Hori \textit{et al.}~\cite{Hori_2006}. While we observe a satisfactory overall agreement with both the experimental data and the 10~K theoretical predictions, the results from Ref.~\cite{Bakalov_2012} gradually diverge from our calculations as the temperature decreases. As demonstrated in our recent study of antiprotonic helium-4 lines~\cite{Jozwiak_2025}, this low-temperature difference can be directly attributed to the semiclassical approximation employed in Ref.~\cite{Bakalov_2012}: below 5~K, the dynamics are governed by only a few partial waves, where the assumption that particles follow classical trajectories naturally breaks down.

\section{Conclusion}
We performed a rigorous theoretical evaluation of collisional effects in three exotic helium atoms: pionic and kaonic helium-4, and antiprotonic helium-3. Utilizing a recent \textit{ab initio} potential energy surface, we first assessed the collisional stability of the metastable states in pionic and kaonic helium. We demonstrated that excited metastable states in pionic helium in the $n=19$ manifold are unstable against a barrierless capture process, which we model using the Langevin framework. These states undergo rapid nuclear capture by the ordinary helium nucleus on picosecond timescales, orders of magnitude faster than their isolated Auger lifetimes. For more tightly bound metastable states across both species, strong short-range repulsion prevents this process. Furthermore, the massive energy gaps to the nearest states ($\sim10^{3}\,\mathrm{K}$) severely suppress inelastic quenching at cryogenic temperatures.

Having established the collisional robustness of these metastable states, we identified the most viable candidate transitions for spectroscopy. Employing rigorous coupled-channel quantum scattering calculations, we evaluated the pressure broadening and pressure shift coefficients for these lines, providing the first comprehensive set of theoretical benchmarks for these exotic systems. We identified a universal physical mechanism governing the collisional effects in all three exotic species: although replacing an electron with a massive particle provides these systems with a molecule-like rovibrational energy structure, their collisional dynamics remain fundamentally atomic. Unlike ordinary molecular systems, where spectral broadening is predominantly driven by collision-induced inelastic transitions, the energy gaps in exotic helium remain vast compared to cryogenic collision energies. This completely suppresses inelastic quenching, leaving the line-shape parameters to be determined almost entirely by elastic dephasing: the accumulation of phase differences between the elastic scattering amplitudes of the initial and final states during the collision.

We anticipate that the results reported here will be highly valuable for the design and interpretation of future high-precision spectroscopic experiments on exotic helium atoms in helium baths. Ultimately, such measurements will be instrumental in testing three-body quantum electrodynamics calculations and refining the fundamental masses of the $\pi^{-}$ and $K^{-}$ mesons.

\section{Data availability}
The cross-sections, rate coefficients, and line-shape parameters computed in this work are provided in the Supplemental Material \cite{SupMat}.

\section{Acknowledgement}
The research is funded by the European Union (ERC-2022-STG, H2TRAP, 101075678). Views and opinions expressed are however those of the author(s) only and do not necessarily reflect those of the European Union or the European Research Council Executive Agency. H.~J.~J. was supported by the National Science Centre in Poland through Project No. 2024/53/N/ST2/02090. 
We gratefully acknowledge the Polish high-performance computing infrastructure PLGrid (HPC Centers: ACK Cyfronet AGH) for providing computer facilities and support within the computational grant, Grant No. PLG/2025/018511. Created out using resources provided by Wroclaw Centre for Networking and Supercomputing (http://wcss.pl).

\appendix

\section{Computational details of coupled-channel quantum scattering calculations}
\label{app:A}
Because coupled-channel (CC) quantum scattering calculations are numerically exact for a given interaction potential, their overall accuracy is determined entirely by the convergence of the specific basis and grid parameters. To ensure sub-percent accuracy across all computed cross-sections, we employed a rigorously tested set of propagation parameters. The radial expansion coefficients of the wave function were propagated using the renormalized Numerov algorithm on an equidistant grid extending from $R_{\mathrm{min}} = 2.5\,a_{0}$ to $R_{\mathrm{max}} = 50\,a_{0}$, with 25 steps per half-de Broglie wavelength. Computations were performed in the total angular momentum representation, which naturally block-diagonalizes the Hamiltonian with respect to the total angular momentum $J$ and spatial parity. To guarantee convergence at the highest collision energies, the calculations incorporated a sufficient number of $J$-blocks, ranging up to $J_{\mathrm{max}} = 60$ for pionic helium, $J_{\mathrm{max}} = 80$ for kaonic helium, and $J_{\mathrm{max}} = 100$ for antiprotonic helium-3.

Because the energy spacing between adjacent states in these exotic atoms is extremely large (on the order of $10^{3}\,\mathrm{K}$), we significantly truncate the rovibrational basis sets. In calculations of inelastic quenching, the basis was strictly limited to the initial metastable state $(v, j)$ and the nearest exothermic de-excitation channel $(v+1, j-1)$. Additional convergence tests revealed that even these nearest neighbors have a negligible impact on the pressure broadening ($\gamma_{0}$) and shift ($\delta_{0}$) coefficients for the selected transitions. Therefore, in our production runs for pionic and kaonic helium, the spectroscopic line-shape calculations were performed utilizing a single-state basis. Specifically, to compute the line-shape parameters for a $(v_{a},j_{a}) \to (v_{b},j_{b})$ transition, two sets of scattering calculations were performed: one containing solely the $(v_{a},j_{a})$ level and the other containing solely the $(v_{b},j_{b})$ level. We emphasize that these are not \textit{single-channel} calculations; each calculation included a certain number of partial waves determined by the triangular rules coupling $j$ and $J$. For antiprotonic helium-3, we adopted the more conservative basis established in our previous work~\cite{Jozwiak_2025}, retaining both the nearest lower and upper neighboring states in the basis set.

The radial expansion coefficients of the interaction potential were evaluated numerically, using Gauss-Legendre quadrature for the angular coordinate ($\theta$) and Simpson's rule for the exotic atom's internal radial coordinate ($r$). In the case of kaonic and pionic helium atoms, the potential expansion was truncated at $\lambda = 11$, while for antiprotonic helium-3, we kept expansion terms up to $\lambda = 7$. The rovibrational wave functions of the exotic atoms, used to evaluate effective expansion coefficients of the PES and coupling matrix elements, were obtained by solving the radial Schr\"{o}dinger equation for the isolated exotic atom on the potential energy curve of Shimamura~\cite{Shimamura_1992}, discretized on an equidistant grid spanning $r=0.15$ to $1.8\,a_{0}$.

\section{Generalized spectroscopic cross-sections}
\label{app:B}
\begin{figure*}[!ht]
\centering\includegraphics[width=0.9\linewidth]{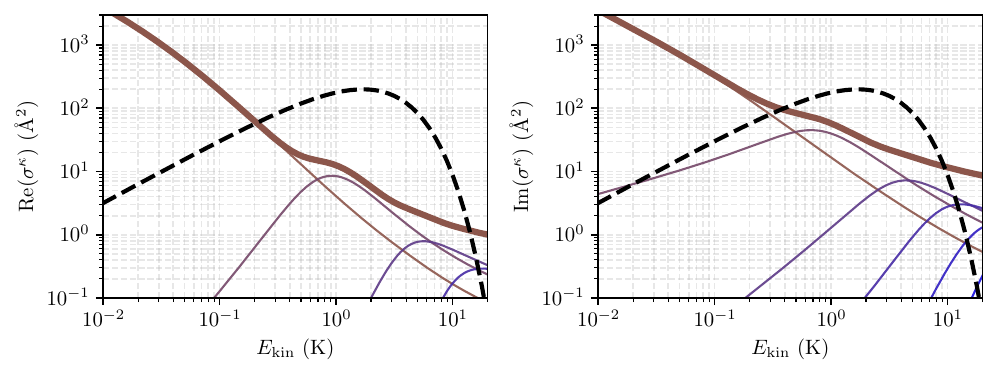}
\caption{Pressure broadening ($\mathrm{Re}(\sigma^{\kappa})$) and pressure shift ($\mathrm{Im}(\sigma^{\kappa})$) cross-sections (thick solid lines) for the selected $(2,27) \to (2,26)$ favored transition in $K^{-\,4}\mathrm{He}^{+}$, along with the partial wave decomposition (thin solid lines) up to $L=6$. The Maxwell--Boltzmann distribution at $T=1.7$~K is shown for reference (dashed black line, arbitrary units).}
\label{fig:gsxs_KHe}
\end{figure*}
In this Appendix, we present selected examples of the computed generalized spectroscopic cross-sections [Eq.~\eqref{eq:gsxs}] to discuss their typical kinetic-energy dependence and partial wave decomposition.

For the vast majority of the transitions evaluated in this work, the generalized spectroscopic cross-sections exhibit a smooth, featureless dependence on kinetic energy, consistent with the behavior previously reported for antiprotonic helium-4~\cite{Jozwiak_2025}. To illustrate this behavior, Fig.~\ref{fig:gsxs_KHe} shows the pressure broadening and shift cross-sections for the $(v=2,j=27) \to (v=2,j=26)$ favored transition in $K^{-\,4}\mathrm{He}^{+}$. As is typical for these systems, the cross-sections monotonically decrease in magnitude at ultracold collision energies, exhibiting only a mild enhancement around 1~K driven by the leading $L=1$ ($p$-wave) contribution. The overall absence of sharp resonant features across these states is a direct consequence of their relatively shallow effective interaction potentials.

\begin{figure*}[!ht]
\centering\includegraphics[width=0.9\linewidth]{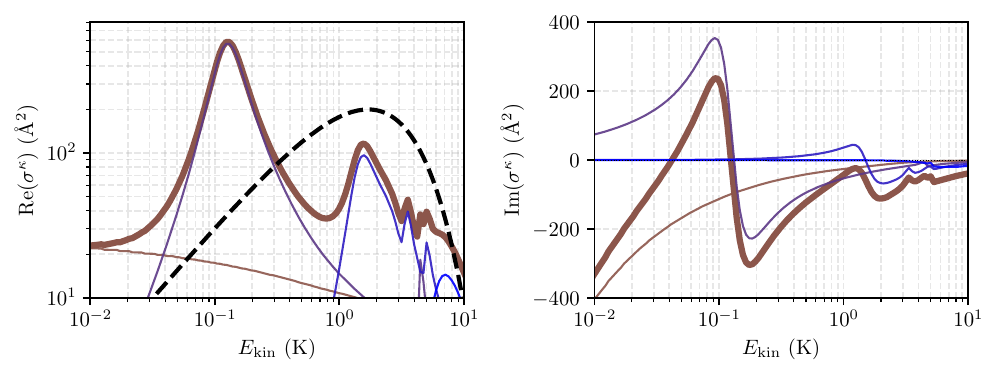}
\caption{Pressure broadening ($\mathrm{Re}(\sigma^{\kappa})$) and pressure shift ($\mathrm{Im}(\sigma^{\kappa})$) cross-sections (thick solid lines) for the selected $(1,16) \to (2,15)$ unfavored transition in $\pi^{-\,4}\mathrm{He}^{+}$, along with the partial wave decomposition (thin solid lines) up to $L=4$. The Maxwell-Boltzmann distribution at $T=1.7$~K is shown for reference (dashed black line, arbitrary units).}
\label{fig:gsxs_piHe}
\end{figure*}

A notable exception to this smooth energy dependence occurs for the metastable $(v=1,j=16)$ state in pionic helium. Fig.~\ref{fig:gsxs_piHe} presents the cross-sections for the unfavored $(v=1,j=16) \to (v=2,j=15)$ transition, which exhibit characteristic scattering resonances. Specifically, we observe distinct shape resonances at collision energies of approximately $0.12$~K and $1.5$~K, which the partial wave decomposition assigns to the $L=1$ ($p$-wave) and $L=2$ ($d$-wave) contributions, respectively. These resonances emerge because the effective isotropic potential for the $(v=1,j=16)$ state [shown as the solid orange line ($n=18$ manifold) in Fig.~\ref{fig:effective_potentials}(a)] is sufficiently deep to support quasi-bound states temporarily trapped behind the angular momentum centrifugal barriers.

We note that these same quasi-bound states manifest as identical resonances in the inelastic quenching cross-sections originating from the $(v=1,j=16)$ state, as previously shown in Fig.~\ref{fig:quench_xs}(a). However, as discussed in the main text, thermal averaging over the Maxwell--Boltzmann distribution effectively washes out these sharp energy-dependent features, and they do not disrupt the monotonic temperature dependence of the rate coefficients, and pressure broadening and pressure shift coefficients presented in Fig.~\ref{fig:PiHe4}.

\bibliography{bibliography}

 \end{document}